%% file: main.tex
\documentclass[runningheads]{llncs}

\usepackage[year=2026]{eccv}
\usepackage{eccvabbrv}

\usepackage{graphicx}
\usepackage{booktabs}
\usepackage{amsmath}
\usepackage{amssymb}
\usepackage{subcaption}
\usepackage{xcolor}
\usepackage{tikz}
\usetikzlibrary{positioning,calc,arrows,backgrounds,fit}
\usepackage[accsupp]{axessibility}

\usepackage[breaklinks,colorlinks,citecolor=eccvblue]{hyperref}

\usepackage{placeins}

\begin{document}

\title{Learning Dense 2D--3D Correspondence for\\ X-ray--to--CT Registration of Knee Bones}
\titlerunning{Dense Correspondence for X-ray--CT Knee Registration}

\author{Rembert Daems\inst{1} \and Jonas Grammens\inst{1} \and Caro Roten\inst{1,2} \\
  Andrew Meyer\inst{1} \and Thomas Luyckx\inst{1,3} \and Matthias Verstraete\inst{1}}
\authorrunning{R. Daems et al.}
\institute{Arthro ID \and KU Leuven \and AZ Delta}

\maketitle

\begin{abstract}
\input{sections/00_abstract}
\end{abstract}

\input{sections/01_intro}
\input{sections/02_related}
\input{sections/03_method}

\input{sections/04_experiments}
\input{sections/05_discussion}
\input{sections/06_conclusion}
\FloatBarrier

\section*{Acknowledgements}
This work was supported by Flanders Innovation \& Entrepreneurship (VLAIO).

\bibliographystyle{splncs04}
\bibliography{main}

\end{document}

%% file: sections/00_abstract.tex
Recovering the 6-DoF pose of the knee bones from a plain radiograph, given the
patient's segmented pre-operative CT, turns a routine low-dose image into a
quantitative measurement of joint geometry, without the added dose of a repeat CT or a
fixed biplanar rig. Classic solutions align a rendered bone silhouette to image edges;
recent alternatives refine pose by backpropagating an image-similarity loss
through a differentiable X-ray renderer. Both operate one patient at a time and are
fragile under a single view. Silhouettes are depth-ambiguous, and differentiable-rendering
refinement has a narrow capture range at substantial per-iteration cost. We instead learn
an \emph{amortized}, subject-agnostic dense 2D--3D correspondence, supervised solely by
projection geometry. One shared-weight model per bone, trained across 758 patients, registers patients unseen
during training. The pose then follows in closed form from a global, initialization-free,
render-free PnP+RANSAC solve. Because
X-ray formation is transmissive, our correspondence target is \emph{transmission-aware}
rather than tied to a single surface. Though trained only to register, the representation
is anatomically semantic: a simple classifier reads a landmark's anatomical region from
its embedding across held-out patients, and the same features separate the
knee's bones into a 2D--3D-consistent identity learned without any bone label.
On a large single-institution cohort the model generalizes well to held-out patients.

%% file: sections/01_intro.tex
\section{Introduction}
\label{sec:intro}

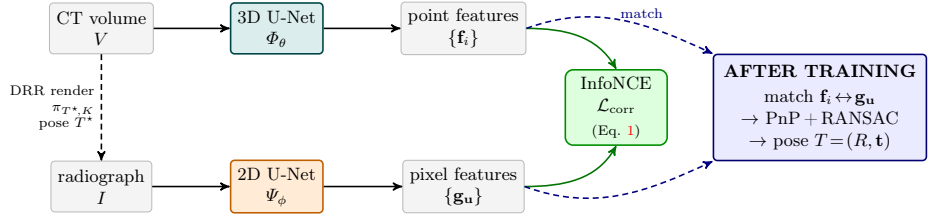
\begin{figure}[t]
\centering
\resizebox{\textwidth}{!}{%
\begin{tikzpicture}[
  font=\small,
  >=stealth',
  box/.style={rounded corners=2pt, draw, minimum height=9mm, align=center, inner sep=3.5pt},
  net3d/.style={box, fill=teal!14, draw=teal!55!black, thick},
  net2d/.style={box, fill=orange!16, draw=orange!75!black, thick},
  data/.style={box, fill=gray!9, draw=gray!55},
  loss/.style={box, fill=green!13, draw=green!55!black, thick, rounded corners=4pt, minimum height=13mm, minimum width=18mm},
  infer/.style={box, fill=blue!8, draw=blue!45!black, thick, align=center, inner sep=5pt, minimum height=13mm},
  flow/.style={->, thick},
  merge/.style={->, thick, green!45!black},
  test/.style={->, thick, densely dashed, blue!50!black},
]
\node[data]  (ct)   at (0,1.4)   {CT volume\\$V$};
\node[net3d] (phi)  at (3.1,1.4)  {3D U-Net\\$\Phi_\theta$};
\node[data]  (femb) at (6.4,1.4)  {point features\\$\{\mathbf{f}_i\}$};
\node[data]  (img)  at (0,-1.4)   {radiograph\\$I$};
\node[net2d] (psi)  at (3.1,-1.4) {2D U-Net\\$\Psi_\phi$};
\node[data]  (gemb) at (6.4,-1.4) {pixel features\\$\{\mathbf{g}_\mathbf{u}\}$};
\node[loss]  (loss) at (9.1,0)    {InfoNCE\\$\mathcal{L}_\text{corr}$\\[1pt]{\scriptsize(Eq.~\ref{eq:infonce})}};
\node[infer] (inf)  at (12.7,0)   {{\footnotesize\textbf{AFTER TRAINING}}\\[2pt]
  match $\mathbf{f}_i\!\leftrightarrow\!\mathbf{g}_\mathbf{u}$\\ $\to$ PnP\,+\,RANSAC\\ $\to$ pose $T\!=\!(R,\mathbf{t})$};
\draw[flow] (ct)  -- (phi);
\draw[flow] (phi) -- (femb);
\draw[flow] (img) -- (psi);
\draw[flow] (psi) -- (gemb);
\draw[flow, densely dashed] (ct.south) -- (img.north)
      node[midway,left,font=\scriptsize,align=right]{DRR render\\$\pi_{T^\star\!,K}$\\pose $T^\star$};
\draw[merge] (femb.east) to[out=0,in=120] (loss.north);
\draw[merge] (gemb.east) to[out=0,in=240] (loss.south);
\draw[test] (femb.east) to[out=20,in=155]
      node[pos=0.62,above,font=\scriptsize]{match} (inf.north west);
\draw[test] (gemb.east) to[out=-20,in=205] (inf.south west);
\end{tikzpicture}%
}
\caption{\textbf{Method overview.} We learn an \emph{amortized} dense 2D--3D
correspondence between a patient's CT and a single radiograph. A 3D U-Net
$\Phi_\theta$ embeds the CT and a 2D U-Net $\Psi_\phi$ embeds the radiograph
into a \emph{shared} feature space. At training time the known pose
$T^\star$ projects each 3D point to its pixel, and a multi-positive InfoNCE loss
(\cref{eq:infonce}) pulls a pixel's feature toward the features of all bone points
along its ray (X-ray is transmissive). Pixel and point features are matched directly
at test time and the $6$-DoF pose $T$
follows from a single global PnP+RANSAC solve that is initialization-free and
non-differentiable in the pose by design.}
\label{fig:overview}
\end{figure}

The six-degrees-of-freedom (6-DoF) pose of the knee bones (femur, patella, tibia, and fibula)
relative to the imaging device is a basic ingredient of in-vivo kinematic analysis and
of a range of pre- and intra-operative planning tasks. Plain radiographs are cheap,
fast, and ubiquitous, and a patient's pre-operative computed tomography (CT) scan, which
automatic tools segment into individual bones~\cite{Wasserthal2023}, is routinely
available. Registering each segmented bone to the radiograph therefore turns
an everyday image into a quantitative, three-dimensional measurement of joint geometry,
with no added dose of a repeat CT and no fixed biplanar rig. We target this problem in
its hardest, most accessible form: given a \emph{single} radiograph and the patient's
CT, recover the 6-DoF pose of a bone.

A single view makes the problem geometrically ill-posed in depth. Motions that
slide the bone along the viewing ray leave its projection almost unchanged, so the
in-plane pose can be recovered to sub-millimetre accuracy while the out-of-plane
component stays weakly constrained \cite{daems2016fluoroscopy,kneevideofluoro2020}.

Prior work falls into two patient-specific families.
First, silhouette or edge
matching \cite{markelj2012review,penney1998comparison} aligns the rendered contour
of the bone to detected image edges. By using only the contour it discards the rich
interior of the radiograph, is depth-ambiguous, and has a narrow basin of
convergence that demands a good initialization. Second, differentiable-rendering
registration \cite{gopalakrishnan2024diffpose} instead backpropagates an
image-similarity loss through a differentiable DRR renderer to refine an initial
pose at test time. Because the renderer is in the optimization loop, iterative
optimization is costly, and gradient-based optimization is prone to local minima.
Both research directions also optimize one patient at a time and learn
no representation that transfers across patients.

We take a different and conceptually simple approach (\cref{fig:overview}). We learn
an \emph{amortized} dense 2D--3D correspondence: a two-encoder model maps every
radiograph pixel and every CT point into a shared embedding, so that a pixel and the
3D point it images receive similar features. Matching pixel features against the patient's CT features yields
dense 2D--3D correspondences, from which pose is recovered in closed form with
Perspective-$n$-Point (PnP) inside a random sample consensus (RANSAC) loop.
This pose estimate is global and initialization-free, and render- and gradient-free
at test time. \emph{The model is
subject-agnostic: a single shared-weight network registers bones and patients it
never saw during training, in contrast to the per-patient optimization of prior work.}

The embedding is learned
\emph{self-supervised} using the projection geometry. This results in a transferable
representation of bone anatomy, for which registration is just one result.
The same features carry a \emph{2D--3D-consistent bone identity}, learned with no bone-level
label, which lets the CT's per-bone segmentation be transferred onto the radiograph. They
also identify anatomical landmarks across patients (\cref{sec:multibone}).

One design issue is specific to X-ray. Unlike opaque surfaces in ordinary
computer-vision pose estimation, a radiograph is transmissive: each pixel integrates
attenuation along its ray, so a pixel corresponds not to one surface point but to all
bone points along that ray. Our correspondence target is therefore
\emph{transmission-aware} (\cref{sec:method}). We train the correspondence from
digitally reconstructed radiographs (DRRs). Since only the 2D encoder consumes a radiograph,
the sim-to-real gap is confined to this sub-network and can be closed by
physics-based rendering or light fine-tuning on paired data, without retraining the
correspondence space (\cref{sec:discussion}).

The closest prior work regresses dense scene coordinates for X-ray-to-CT registration
on the pelvis \cite{shrestha2023xray}. That method regresses the absolute 3D
coordinates of one specific CT and is trained per patient. It learns no transferable
representation. Our correspondence is an amortized embedding
that transfers to unseen patients.

\paragraph{Contributions.}
\begin{itemize}
  \item We propose an \textbf{amortized, subject-agnostic dense 2D--3D correspondence}
    for single-view X-ray-to-CT registration of the knee bones: a single shared-weight
    model maps radiograph pixels and CT points into a common embedding and registers
    held-out patients without per-patient training or initialization.
  \item We introduce a \textbf{transmission-aware correspondence target}: because X-ray
    is additive along the ray, we supervise all points along a pixel's ray with a
    multi-positive InfoNCE loss, rather than a single opaque-surface coordinate.
  \item We show the representation is \textbf{reusable beyond pose}: trained only to
    register, it develops an \textbf{emergent, label-free bone identity} that is consistent
    across the 2D and 3D towers (a pixel's nearest 3D match lands on the correct bone
    $96\%$ of the time on held-out patients), enough to transfer the
    CT's segmentation onto the radiograph and to identify knee landmarks across patients
    ($98\%$ region accuracy, leave-one-patient-out). This is evidence of a general
    anatomical embedding rather than a pose-specific feature (\cref{sec:multibone}).
  \item We \textbf{benchmark against differentiable-rendering registration} on capture
    range and cost, distinguishing patient-specific \emph{test-time} differentiable
    rendering (DiffDRR/DiffPose) from the \emph{training-time} rendering we use:
    differentiable rendering is more accurate only within a narrow basin near the ground
    truth, whereas our initialization-free solve is globally robust at two orders of
    magnitude lower cost.
  \item We \textbf{analyze the single-view limit}, decomposing the error into in-plane and
    depth components to show that correspondences are essentially solved and the residual
    is geometric depth ambiguity, which a second calibrated view resolves.
\end{itemize}

%% file: sections/02_related.tex
\section{Related Work}
\label{sec:related}

\subsection{2D--3D X-ray-to-CT registration}
\textbf{Intensity- and feature-based registration.}
Classical 2D--3D registration follows a standard taxonomy \cite{markelj2012review}:
intensity-based methods optimize a similarity between a rendered DRR and the
radiograph (normalized cross-correlation, mutual information, or gradient
correlation \cite{penney1998comparison,desilva2016similarity,otake2012multiview}),
while feature-based methods match geometric primitives such as contours. Both
rely on iterative optimization with a small capture range and strong sensitivity to
initialization, and are assessed with the standardized gold-standard protocol
(mean target registration error, capture range, success rate) that we adopt
\cite{vandekraats2005standardized}; the move toward learning-based components is
surveyed in \cite{unberath2021impact}. Throughout, registration is solved one
patient at a time, with no representation shared across patients.

\subsection{Silhouette- and model-based knee registration}
\textbf{Single-plane fluoroscopy.}
Our baseline lineage is model-based single-plane registration of the knee. Founding
work established silhouette-to-contour alignment \cite{banks1996accurate}; a
theoretical-accuracy analysis for natural bones isolates edge-detection bias as the
limiting factor and reports larger patellofemoral than tibiofemoral error
\cite{fregly2005theoretical}; and subsequent methods contrast silhouette against
direct-intensity matching \cite{mahfouz2003robust}, automate contour detection
\cite{prins2011integrated}, and underpin model-based Roentgen stereophotogrammetry
\cite{kaptein2003new}. The single-plane setup is accurate in-plane but weak
out-of-plane: validated against a motion-capture reference, it attains
sub-millimetre and sub-degree in-plane accuracy yet markedly larger depth error
\cite{daems2016fluoroscopy,kneevideofluoro2020}. Modern deep baselines still segment
and match a silhouette \cite{jensen2023jointtrack,vogl2022personalised}. This
recurring weakness, depth ambiguity and the patella in particular, motivates
replacing the contour with a dense interior signal.

\subsection{Learned correspondence and differentiable rendering for X-ray}
\textbf{Differentiable rendering.}
DiffPose \cite{gopalakrishnan2024diffpose} regresses an initial pose and refines it
at test time by backpropagating an image-similarity loss through a differentiable
DRR renderer; follow-up work trains rapid patient-specific networks for the same
optimization loop \cite{gopalakrishnan2025xvr}. The approach is powerful but, as we
argue in \cref{sec:critique}, patient-specific and fragile: a narrow capture range
and a renderer kept inside the test-time loop.
\textbf{Scene-coordinate regression.}
Closest to us, \cite{shrestha2023xray} regresses dense scene coordinates of one
pelvis CT and solves pose with PnP; it is trained per patient and learns no
transferable embedding. Sparse learned landmarks likewise drive 2D/3D registration
\cite{grupp2020automatic,bier2018xray}, and ray-embedding subspaces enable arbitrary
landmark detection in X-ray images \cite{shrestha2024rayemb}. We differ by learning
an \emph{amortized}, transmission-aware correspondence embedding that transfers across
patients rather than being fit per patient.

\subsection{Dense 2D--3D correspondence for pose in computer vision}
\textbf{Correspondence-to-pose.}
Our formulation adapts the dense-correspondence-then-PnP paradigm from object pose
estimation. NOCS regresses a normalized object-coordinate map \cite{wang2019nocs};
EPOS handles symmetric objects with surface fragments \cite{hodan2020epos}; and
SurfEmb learns continuous surface embeddings with a contrastive (InfoNCE) objective
\cite{haugaard2022surfemb}, which our loss most closely follows.
\textbf{Render-and-compare.}
An alternative line iteratively optimizes pose by rendering the object and comparing it
against the observation at test time: deep iterative matching \cite{li2018deepim},
feature-metric refinement through a learned renderer \cite{iwase2021repose},
multi-view \cite{labbe2020cosypose} and novel-object \cite{labbe2022megapose}
render-and-compare, or matching against learned intermediate representations such as
center and curvature heatmaps \cite{deroovere2025ccpose}. These refiners evaluate a
network (or a renderer) at \emph{every} optimization iteration; we deliberately take the
correspondence-to-PnP route to avoid any test-time optimization and its repeated
per-iteration cost, and adapt it to transmission imaging, where a pixel maps to
\emph{all} points along its ray rather than to a single opaque surface point.

%% file: sections/03_method.tex
\section{Method}
\label{sec:method}

\subsection{Overview and notation}
\label{sec:formulation}
We register each bone independently. Let $V$ denote a patient-specific bone
(its segmented CT density volume), and let
$\mathcal{P}=\{\mathbf{X}_i\in\mathbb{R}^3\}_{i=1}^{N}$ be points sampled from its
intensity-weighted density (our method works directly on the volume, not on derived surfaces). Given a single
calibrated radiograph $I$ with known intrinsics $K$
(cone-beam source--detector geometry), we seek the rigid pose
$T=(R,\mathbf{t})\in SE(3)$ that places $V$ in the camera frame, so that the
projection $\pi_{T,K}(\mathbf{X})$ of each point lands on its true image
location. We cast this as \emph{dense 2D--3D correspondence followed by a
geometric solve}: predict, for points on the bone and pixels in the image, a
shared embedding in which true correspondences match; then recover $T$ with PnP. We obtain
$T$ from a \emph{global} correspondence solve, \emph{never} by optimizing pose through a
renderer (\cref{sec:inference}).

\subsection{Two-encoder architecture}
\label{sec:encoders}
We use two encoders mapping to a common $d$-dimensional space, in the spirit of
learned surface embeddings for object pose~\cite{haugaard2022surfemb,wang2019nocs}.
A \emph{3D encoder} $\Phi_\theta$ embeds each point conditioned on the patient
bone, $\mathbf{f}_i=\Phi_\theta(V)(\mathbf{X}_i)\in\mathbb{R}^d$. It is a $3$D
U-Net over the CT density volume $V$ producing a feature volume, from which
per-point features are read by trilinear sampling. It operates on local density
\emph{content}, i.e., the bone geometry around each point, rather than on coordinates
directly, which would only encode absolute position. This is what makes the features
semantic and transferable across subjects. A \emph{2D
encoder} $\Psi_\phi$ maps the
radiograph to a dense feature map, $\mathbf{g}_{\mathbf{u}}=\Psi_\phi(I)[\mathbf{u}]
\in\mathbb{R}^d$ at pixel $\mathbf{u}$. The two encoders are trained jointly so that
a 3D point and its true projection are nearest neighbours in feature space.
Because the 3D encoder reads the CT directly, the model is \emph{subject-agnostic}:
a new patient is registered without any per-patient training, unlike
patient-specific differentiable-rendering pipelines~\cite{gopalakrishnan2024diffpose}.

\subsection{Correspondence under projective transmission geometry}
\label{sec:transmission}
At training time the known pose $T^\star$ defines, for each 3D point
$\mathbf{X}_i$, its ground-truth image location $\mathbf{u}_i=\pi_{T^\star,K}(\mathbf{X}_i)$,
giving free dense supervision from synthetic radiographs (\cref{sec:data}).
Unlike opaque-surface pose estimation, radiographs are \emph{transmissive}: with
no occlusion, \emph{every} 3D point along a ray projects to the same pixel, so the
pixel$\rightarrow$3D map is one-to-many (near and far cortex) and a pixel's
correspondence target is the whole \emph{set} of points on its ray, not a single
surface point. This corresponds to the physics of image formation, where the radiograph
intensity is the additive line integral along the ray, and we adopt the same
additive view for features: a pixel's positives are all points that project onto
it (\cref{sec:losses}). The dense cortices (ray entry and exit) dominate the set. The
correspondence is also mesh-free, read directly off the projected points with no
isosurface to choose, unlike opaque-object correspondence, where each pixel maps to one
visible surface point~\cite{haugaard2022surfemb,hodan2020epos}.

\subsection{Training loss}
\label{sec:losses}
Under transmission a pixel $\mathbf{u}$ has a \emph{set} of positives
$\mathcal{P}(\mathbf{u})=\{i:\pi_{T^\star,K}(\mathbf{X}_i)=\mathbf{u}\}$, i.e., the
sampled points projecting onto it. We use a multi-positive InfoNCE whose softmax
numerator \emph{sums} over that set (the additive, log-sum-exp analog of the line
integral that forms the image),
\begin{equation}
  \mathcal{L}_\text{corr}
  = -\sum_{\mathbf{u}}\log
  \frac{\sum_{i\in\mathcal{P}(\mathbf{u})}\exp\!\big(\langle \mathbf{g}_{\mathbf{u}},\mathbf{f}_i\rangle/\tau\big)}
       {\sum_{j}\exp\!\big(\langle \mathbf{g}_{\mathbf{u}},\mathbf{f}_j\rangle/\tau\big)},
  \label{eq:infonce}
\end{equation}
with temperature $\tau$ and the denominator over the point key bank. This keeps a
proper (multimodal) per-pixel distribution over the bone rather than forcing a single match. A simpler variant that
\emph{sums} the positives' feature vectors and matches the aggregate is consistently worse
($0.96$ vs.\ $0.98$ correspondence accuracy on a real femur): the sum saturates the
embedding dimension, which the log-sum-exp form avoids.

\subsection{Inference: matching and PnP}
\label{sec:inference}
At test time the pose is unknown, so we do \emph{not} project. We match the
patient's 3D embeddings against the radiograph feature map directly in feature
space, yielding a set of candidate 2D--3D correspondences, and solve for $T$ with
PnP inside RANSAC. This is a single global solve requiring no initialization,
test-time rendering or gradients through pose.
This contrasts with prior literature based on pose estimation by gradient descent through
a renderer over a non-convex objective with a narrow capture
range~\cite{gopalakrishnan2024diffpose,gao2020prost}. We test robustness
to initialization in \cref{sec:capture} and discuss the differences with our method in \cref{sec:critique}.

%% file: sections/04_experiments.tex
\section{Experiments}
\label{sec:exp}
We evaluate on real patient femurs:
cross-patient generalization, the central subject-agnostic claim (\cref{sec:crosspatient});
robustness to initialization against a differentiable-rendering baseline
(\cref{sec:capture}); validation on real plain radiographs (\cref{sec:real}); and an
exploratory extension to whole-knee multi-bone identity (\cref{sec:multibone}).

\subsection{Data and DRR synthesis}
\label{sec:data}
Real femurs are segmented by an in-house automatic tool from a large
single-institution cohort of $766$ knee CTs, which we split by
patient into $758$ for training and $8$ held out for testing; the split is by identity,
so no test patient appears in training. Hounsfield units are mapped to
attenuation on an isotropic, centered grid, so the 3D encoder reads genuine cortical and
trabecular content. We render $128\times128$ DRRs (volume ray-marches) at known poses; a
known pose labels every pixel--point pair, giving free, geometrically exact dense
supervision. Poses span the full clinical range: the view sweeps $\pm90^\circ$ in azimuth about the
bone's long axis, from the canonical anterior--posterior (AP) view through oblique to
lateral, with a further $\pm15^\circ$ out-of-axis tilt and $\pm20$\,mm in-plane translation.
A single model trained over this range matches both the AP and the lateral view, so the same
weights serve the single- and two-view solves (\cref{sec:crosspatient}).
We further evaluate on an in-house cohort with paired \emph{real} radiographs
and annotator-validated poses (\cref{sec:real}). To our knowledge no \emph{public}
dataset pairs patient knee CT with real plain radiographs for registration: the
Osteoarthritis Initiative (OAI)~\cite{peterfy2008oai} provides
unpaired knee X-ray and MRI, and public 2D/3D registration benchmarks target the
hip/pelvis~\cite{grupp2020automatic} or use knee-implant phantoms rather than patient
knees. This motivates our in-house acquisition.

\subsection{Evaluation protocol and metrics}
\label{sec:metrics}
We follow the standardized 2D--3D registration protocol
\cite{vandekraats2005standardized}. The primary metric is the mean Target Registration
Error (mTRE), the mean distance between bone points under the estimated and
ground-truth pose. We report its median over poses and the success rate (SR), the
fraction of poses below $5$\,mm. Since a single view is depth-degenerate, we also
decompose the camera-frame error into an \emph{in-plane} (detector) and an
\emph{out-of-plane} (depth) component, which separates correspondence quality from
depth ambiguity. Correspondence accuracy is the fraction of matched points
within $3$\,px of their pixel (mesh-free, the only ground truth
without a surface). We also report the \emph{capture range}, the
initialization offset up to which registration still succeeds (\cref{sec:capture}).

\subsection{Cross-patient generalization}
\label{sec:crosspatient}
\begin{figure}[t]
\centering
\includegraphics[width=\textwidth]{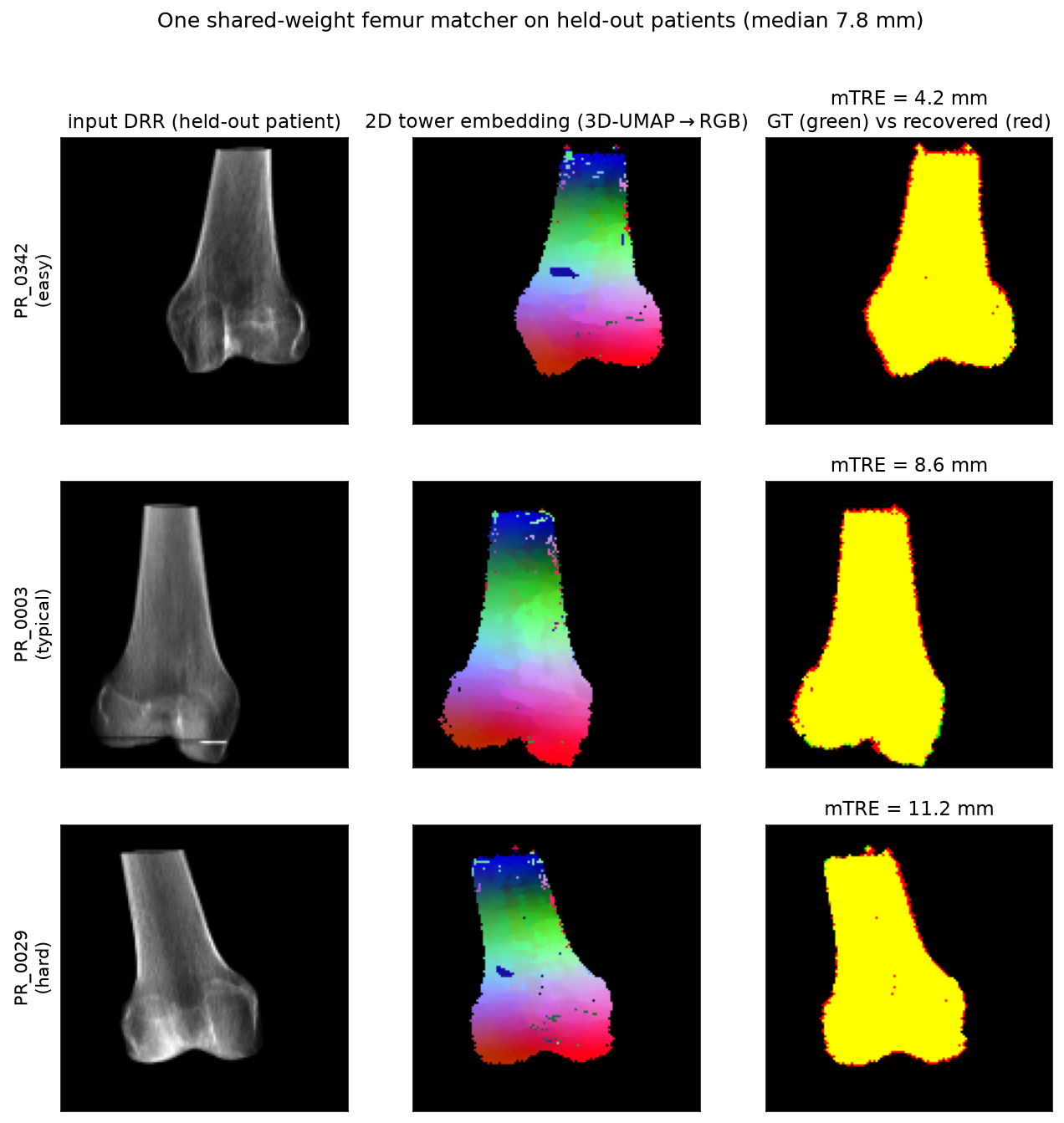}
\caption{Cross-patient embedding transfer by one shared-weight, mesh-free model on
three \emph{held-out patients} (easy/typical/hard). Left: the input DRR; middle: the
$2$D-encoder embedding (3D-UMAP$\rightarrow$RGB, masked to the femur); right: the
recovered-pose silhouette (red) over ground truth (green). On bones never seen in
training the embedding is smooth, informative and anatomically coherent. This shows
that the encoder keys
on transferable local content rather than memorized position. The hard case is the long,
near-cylindrical shaft, where single-view depth is least observable.}
\label{fig:crosspatient}
\end{figure}

We train a shared-weight, mesh-free model on $758$ patient
femurs and test on $8$ \emph{held-out patients} (disjoint identities, no
per-patient training). \Cref{tab:crosspatient} reports pose recovery on seen and
unseen patients. On unseen patients the learned $2$D embeddings stay smooth
and anatomically coherent (\cref{fig:crosspatient}), and we observe \emph{no} significant
generalization gap: the model registers unseen anatomy as well as seen.

\paragraph{Decomposition into in-plane and depth error.} We decompose the camera-frame
error into an in-plane (detector) and an out-of-plane (depth) component
(\cref{tab:crosspatient}). For both the femur and the patella the in-plane component is
sub-millimetre while the depth component dominates the residual: the dense correspondences
are excellent, and the large $\text{mTRE}$ is almost entirely the depth that a single
radiograph cannot observe. This residual barely moves with model size or training budget:
capacity changes how often the solve succeeds but not the residual itself, which points to a
geometric limit rather than a limit of capacity. What removes it is a second view.

\paragraph{Two views.} Since the residual is out-of-plane,
an orthogonal second radiograph should remove it. Our single-view model already spans the
wide clinical pose range ($\pm90^\circ$), so it matches an AP \emph{and} a
lateral view. We fuse both views' correspondences in a single calibrated multi-view PnP
solve, with no per-view retraining. The second view reduces the depth component, which
lowers the median $\text{mTRE}$ and raises the success rate (\cref{tab:crosspatient}).

\begin{figure}[t]
\centering
\includegraphics[width=\textwidth]{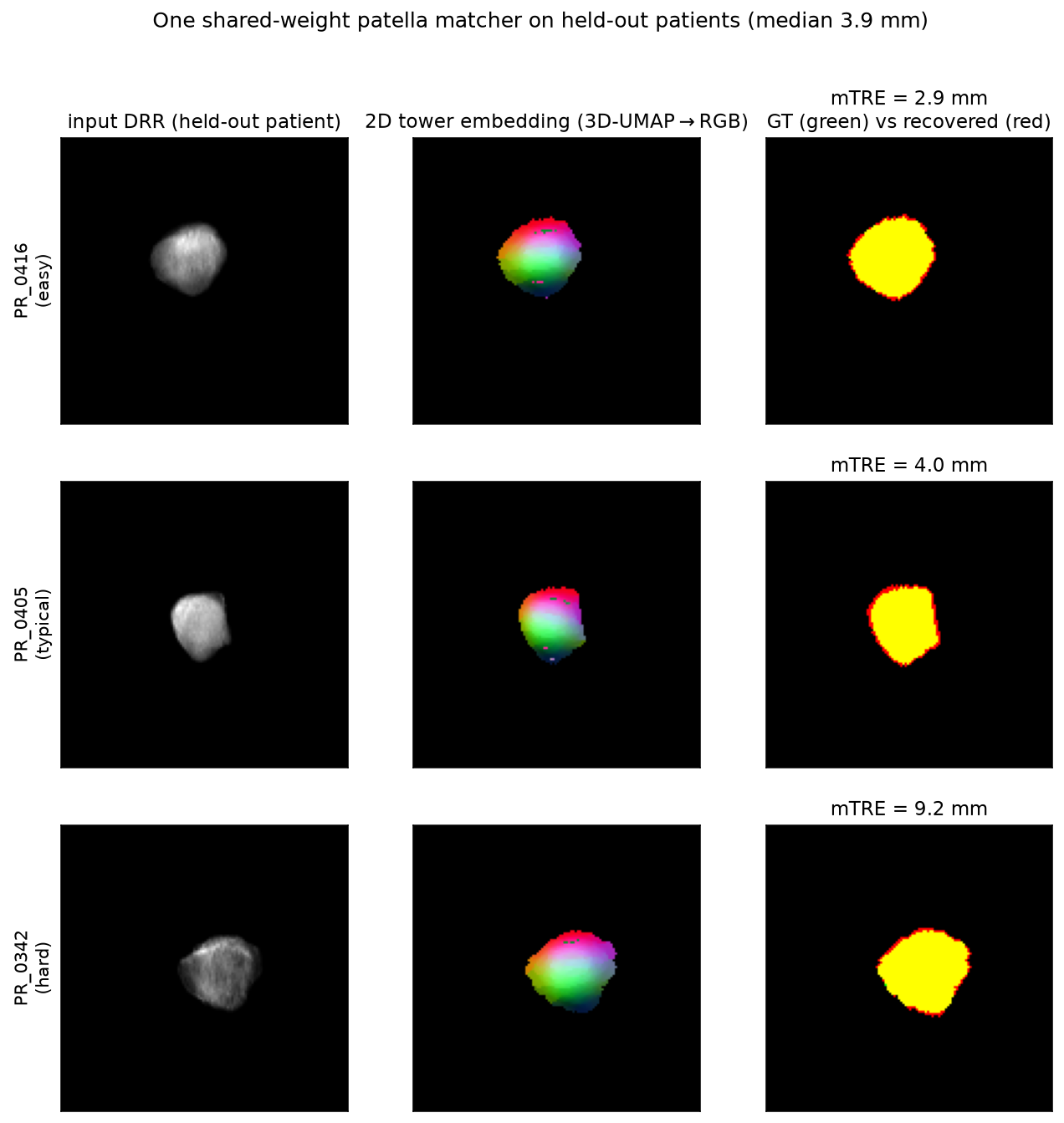}
\caption{The same shared-weight, mesh-free recipe on the \emph{patella}, three
\emph{held-out patients} (easy/typical/hard). Left: the input DRR; middle: the
$2$D-encoder embedding (3D-UMAP$\rightarrow$RGB, masked to the patella); right: the
recovered-pose silhouette (red) over ground truth (green). The embedding is smooth and
consistent across patients on this near-circular bone that has no usable contour, and the
recovered silhouette overlaps ground truth almost exactly in plane; the residual is again
single-view depth. Median $\text{mTRE}$ over the $8$ held-out patients is $4.7$\,mm
(\cref{tab:crosspatient}).}
\label{fig:patella}
\end{figure}

\paragraph{Patella.} We train
one shared-weight model across patients on the patella in isolation, and test on held-out
patients at the clinical pose range (\cref{tab:crosspatient,fig:patella}). It reaches $0.99$
correspondence accuracy with essentially no seen-to-unseen gap, despite the near-circular
silhouette, and as with the femur its accuracy is limited by single-view depth.

\begin{table}[t]
\centering
\caption{Cross-patient generalization, one shared-weight model
\emph{per bone}, $758$ train, $8$ held-out patients, held-out poses. mTRE, in-plane and
depth are medians, where in-plane and depth are the detector-plane and depth components of the
camera-frame error, and rotation and translation are means over successful solves. Seen and
unseen patients are statistically indistinguishable, indicating no generalization gap.
The residual is mostly out-of-plane depth which is largely resolved with a second calibrated view.}
\label{tab:crosspatient}
\resizebox{\textwidth}{!}{%
\begin{tabular}{lccccccc}
\toprule
 & Corr.\ acc.\ $(<\!3\,\text{px})$ & mTRE (mm) & In-plane (mm) & Depth (mm) & Rot.\ $(^\circ)$ & Trans.\ (mm) & SR $(<\!5\,\text{mm})$ \\
\midrule
\multicolumn{8}{l}{\emph{Femur}} \\
\quad Seen patients               & $0.83$ & $8.1$ & $1.7$ & $7.8$ & $7.1$ & $11.7$ & $30\%$ \\
\quad \emph{Unseen}, single view  & $0.85$ & $6.5$ & $1.3$ & $6.3$ & $5.7$ & $8.6$ & $38\%$ \\
\quad \emph{Unseen}, two views    & $0.84$ & $\mathbf{2.4}$ & $1.1$ & $\mathbf{1.8}$ & $4.2$ & $4.2$ & $\mathbf{75\%}$ \\
\midrule
\multicolumn{8}{l}{\emph{Patella}} \\
\quad Seen patients               & $0.99$ & $6.5$ & $0.7$ & $6.4$ & $6.5$ & $7.9$ & $38\%$ \\
\quad \emph{Unseen} patients      & $0.99$ & $\mathbf{4.7}$ & $0.7$ & $4.4$ & $6.5$ & $5.6$ & $\mathbf{53\%}$ \\
\bottomrule
\end{tabular}%
}
\end{table}

\subsection{Robustness to initialization: capture range vs.\ differentiable rendering}
\label{sec:capture}
Our solve is global and initialization-free, while the differentiable-rendering alternative
refines a pose by gradient descent through a renderer and so depends on its
initialization. On the held-out patients we seed a DiffDRR~\cite{gopalakrishnan2022diffdrr}
optimizer following DiffPose~\cite{gopalakrishnan2024diffpose} (NCC loss, Adam) at the
ground truth plus an offset $\Delta$ in rotation and translation, and sweep $\Delta$. Our
solve ignores the seed, so its success rate is flat. The baseline's target is rendered by
DiffDRR itself, so at $\Delta=0$ it recovers the pose to $\approx\!0$\,mm, a fair comparison.

\begin{figure}[t]
\centering
\includegraphics[width=\textwidth]{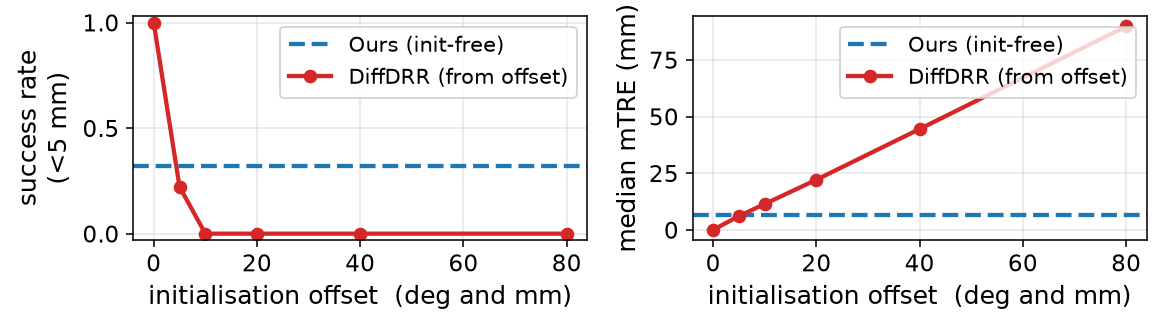}
\caption{\textbf{Capture range and cost.} Success rate (left) and median $\text{mTRE}$
(right) vs.\ initialization offset $\Delta$ (deg and mm), for differentiable-rendering
registration (DiffDRR, red) and our init-free solve (blue, dashed). DiffDRR is most accurate at $\Delta\!=\!0$ but its success collapses within a few
deg/mm; ours dominates beyond the crossover, at ${\sim}315\times$ lower per-registration
cost. $6$ held-out patients $\times\,12$ poses.}
\label{fig:capture}
\end{figure}

\Cref{fig:capture} shows the result. From a near-perfect initialization the
differentiable-rendering solve is the \emph{most accurate} method, with $100\%$ success at
$\Delta=0$, below our depth-limited floor. Its capture range is however narrow: success
drops to $19\%$ at $5$ and to $\approx\!0\%$ by $10$ (deg/mm), crossing below our
init-free solve within a few degrees. Ours is flat for \emph{any} offset, at $6$\,ms per
registration against $1.9$\,s for the optimization, some $315\times$ cheaper. Differentiable
rendering thus reaches sub-pixel accuracy inside a narrow basin at a high cost, while a
global correspondence solve trades peak accuracy for initialization-free robustness.

\subsection{Real radiographs}
\label{sec:real}
Our external test is on \emph{real} plain radiographs. We assemble a cohort of knees whose
segmented pre-operative CT we hold, each with per-view ground-truth femur poses from an
in-house tool in which an operator aligns the CT to every radiograph ($26$ patients, $49$
views, AP and lateral). We evaluate on \emph{held-out} patients with the synthetic-study
protocol (median mTRE, success rate, in-plane/depth decomposition). Projecting each CT at
its pose overlays the femur in every view, confirming the paired data and the calibration.

Only the $2$D encoder ever consumes a radiograph, so the sim-to-real gap is confined to it.
We narrow it by domain-randomizing its input during training (soft-tissue background,
gamma/contrast, blur, noise), leaving the correspondence geometry untouched. We present
the real films as a \emph{feasibility study}. The appearance transfers: the $2$D encoder
produces a smooth, anatomically coherent embedding directly on a real film. Optimizing the
learned-feature cost in pose, a feature-space registration without a per-patient renderer,
recovers the in-plane pose with a \emph{wide capture range}, flat to about $30^\circ$ of
initialization error, while an intensity (DiffDRR) cost on the same films is accurate only
in a narrow basin. This reproduces the capture-range result of \cref{sec:capture} on real
data. Fine-tuning \emph{only} the $2$D encoder on the real films, with the correspondence
space frozen, cuts the held-out in-plane error from a median $9.4$ to $4.6$\,mm. Single-view
accuracy is not yet clinical: the in-plane error is of order $10$\,mm, depth stays
unobservable from one view, and discrete correspondence$+$PnP, robust on synthetic data, is
brittle on one real planar view. A larger annotated cohort, stronger encoder adaptation, and
the second view that removes depth are the main directions to close it.

\paragraph{The effect of overlying tissue.} A controlled synthetic study isolates the effect
of overlying anatomy. We re-render every DRR from the \emph{full} CT, so the femur sits in
the patient's real soft tissue and adjacent bones, and leave the $3$D encoder and the
correspondence ground truth unchanged. At the wide pose range the clutter roughly halves the
confident correspondences on held-out patients, giving a correspondence accuracy of $0.48$
(vs.\ $0.83$ bone-only). The median $\text{mTRE}$ is nevertheless essentially unchanged at
$6.8$ (vs.\ $7.4$)\,mm: the surviving correspondences still constrain PnP, so the $2$D
encoder keys on bone through the tissue. The residual real-data gap is therefore dominated
by $2$D bone \emph{localization}, the step the fine-tune addresses, and not by a collapse of
the embedding.

\subsection{Emergent semantic structure}
\label{sec:multibone}
The self-learning objective is agnostic to anatomical structures. Because it is supervised
by projection geometry, however, the embedding ends up encoding local \emph{bone geometry},
consistent across patients. We read this out in
two ways. The embedding separates the knee into its \emph{bones}, and within a bone it
identifies named \emph{landmarks}.

\begin{figure}[t]
\centering
\includegraphics[width=0.86\textwidth]{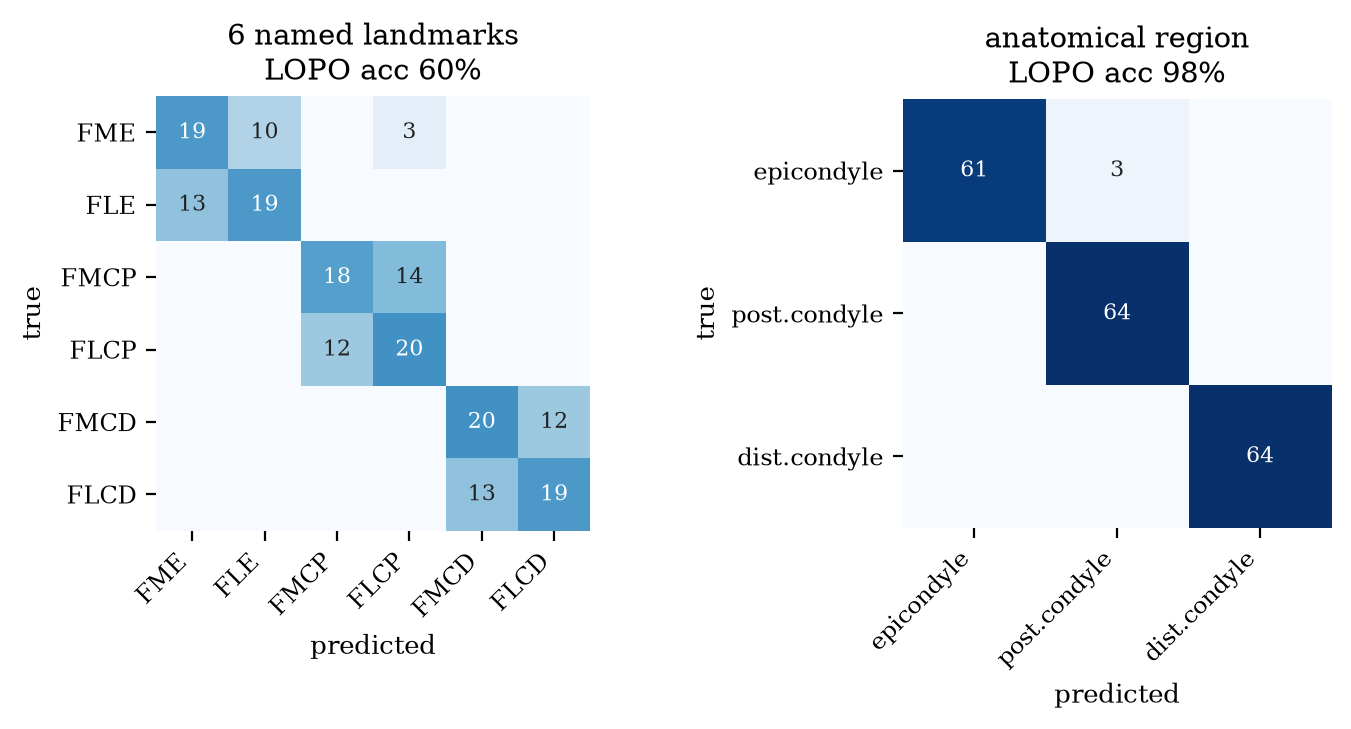}
\caption{\textbf{The registration-only embedding is anatomically semantic.} A
nearest-centroid classifier recovers a knee landmark's identity from its $3$D-tower
embedding alone, leave-one-\emph{patient}-out ($32$ patients, $6$ femoral landmarks).
Left: the $6$-class confusion is block-diagonal, where it is mostly medial\,$\leftrightarrow$\,lateral
within an anatomical region (the
expected left/right mirror ambiguity of a chirality-agnostic matcher). Right: collapsing
to anatomical region gives $98\%$, evidence that the embedding encodes a
shared anatomical code rather than a per-patient position.}
\label{fig:landmark}
\end{figure}

\paragraph{Decoding anatomical landmarks across patients.}
Using an in-house tool that marks standard knee landmarks on the CT (femoral epicondyles
and the medial/lateral posterior and distal condyles), we sample the $3$D encoder's
embedding at each landmark and ask whether its identity is recoverable from that vector
alone, across patients. A nearest-centroid classifier, strictly
leave-one-\emph{patient}-out, resolves the anatomical \emph{region} (epicondyle vs.\
posterior vs.\ distal condyle) at $98\%$ (\cref{fig:landmark}). Most errors are medial\,$\leftrightarrow$\,lateral
\emph{within} a region, because the models are trained chirality-agnostic and
the sides are near-reflections. The embedding is
thus a shared code for bone geometry rather than a per-patient position memory: a landmark
``looks like'' the same landmark in patients never seen. Note that this probes the frozen
representation and is not a trained detector, and it works best where the local shape is
distinctive, as in the femur.

\begin{figure}[t]
\centering
\includegraphics[width=\textwidth]{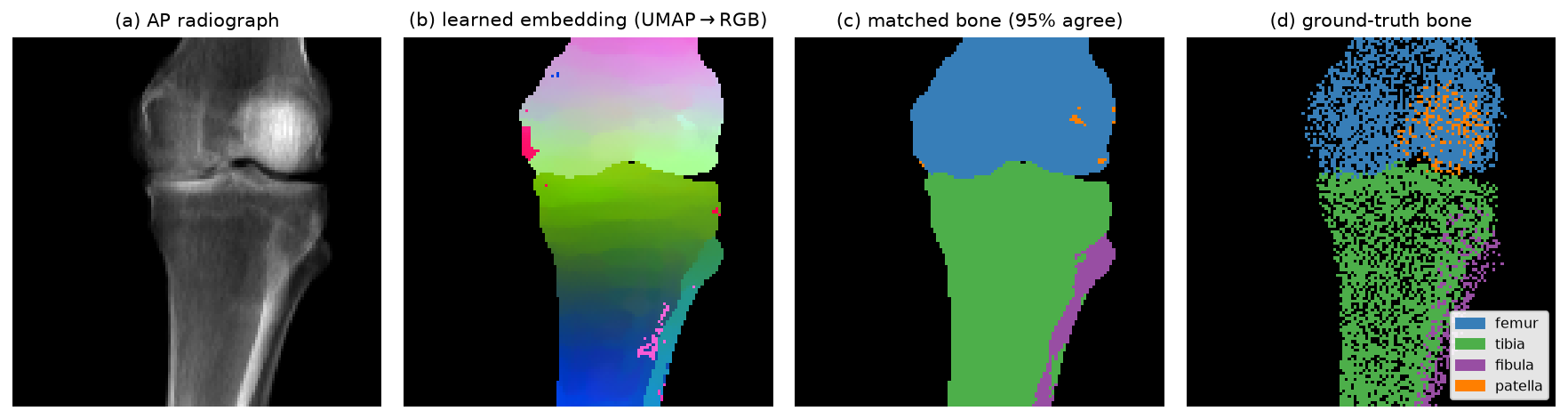}
\caption{\textbf{Emergent, label-free bone identity.} The
whole-knee model is trained with no bone-identity label, only 2D--3D dense correspondence.
(a) the AP radiograph; (b) the learned 2D-encoder embedding (3D-UMAP$\to$RGB), a continuous,
label-free feature map; (c) each foreground pixel coloured by the bone of its nearest CT
match, i.e., the CT's segmentation transferred through the correspondence; (d) the
ground-truth bone map (projected labelled points). The discrete bone routing in (c)
falls out of the continuous embedding in (b): the held-out bones agree with ground
truth for $96\%$ of single-bone pixels, on average. Thus, a correspondence-only embedding separates the knee's bones and lets their
CT segmentation be read off the film.}
\label{fig:segmentation}
\end{figure}

\paragraph{Emergent bone identity.}
We feed the matcher the \emph{entire} knee (femur, tibia, fibula and patella in one
density volume), without ever telling it there is more than one bone. Training
follows the same mesh-free, projection-based recipe (\cref{sec:transmission,sec:losses}),
with no bone-identity label ($150$ train, $8$ held-out patients). Bone identity
emerges unsupervised: to give every pixel a distinguishable $3$D match the embedding
must separate the bones, and it does so consistently across the $2$D and $3$D encoders. We
quantify this with the \emph{2D--3D bone-identity consistency}, the fraction of single-bone
pixels whose nearest $3$D match lies on the same bone, which reaches $96\%$ on held-out
patients. This lets us read a per-bone map off the radiograph by \emph{transferring the
CT's segmentation through the correspondence}: colouring each pixel by its matched bone
recovers femur, tibia, fibula and patella (\cref{fig:segmentation}).

%% file: sections/05_discussion.tex
\section{Discussion and Limitations}
\label{sec:discussion}

\subsection{Our method vs. differentiable rendering}
\label{sec:critique}
Given a good
initialization (e.g.\ intraoperative tracking seeded by the previous frame) and
well-calibrated intensities, gradient descent through a differentiable renderer is
accurate and, with a patient-specific network, fast
\cite{gopalakrishnan2024diffpose,gopalakrishnan2025xvr}. The difficulty is the cold start:
a non-convex landscape with a narrow capture range, a renderer cost per iteration, and a
network trained per patient. Our correspondence-then-PnP solve is complementary (global,
initialization-free, render-free, amortized), at the price of spending its accuracy budget
on correspondences and inheriting the single-view depth limit below. The two compose,
since our global solve is a natural initializer for a differentiable refinement.

\subsection{Limitations}
\paragraph{Single-view depth.}
A single radiograph cannot observe depth: the residual is almost entirely
out-of-plane while the in-plane component is sub-$2$\,mm (\cref{sec:crosspatient}). The
degeneracy is geometric, not a modeling failure, and a calibrated second view removes most
of it. The same geometry limits \emph{orientation} under full $SO(3)$: a near-symmetric
bone projects almost identically from distinct orientations. A model-free probe makes this
concrete: for $84\%$ of orientations some pose more than $90^\circ$ away yields a DRR at
least as image-similar as the closest $<\!30^\circ$ neighbour. Single-view
results should thus be read as in-plane correspondence with a known, view-induced depth
(and, under wide poses, orientation) uncertainty.

\paragraph{Sim-to-real gap.}
We train on DRRs, which omit scatter, beam hardening, and soft tissue. Because only the 2D
encoder consumes a radiograph, this gap is confined to a single sub-network, closable
without retraining the correspondence space (\cref{sec:real}). Yet, single-view
accuracy on real films is not yet clinical, we present it as a feasibility study.

\paragraph{Cohort, segmentation, and joint structure.}
With a large training cohort the seen-to-unseen gap is small, and the residual error is
single-view depth, not cohort size or capacity (\cref{sec:crosspatient}). The method
assumes an accurate CT segmentation of each bone and registers the femur, patella, tibia,
and fibula independently, ignoring the joint's kinematics. The patella is the hardest case,
being small, mobile and low-contrast. Its faint outline is itself hard to extract on a real
film, which is where an appearance-based dense signal helps most over a contour that must
first be found \cite{fregly2005theoretical}.

\paragraph{Baselines and beyond pose.}
Silhouette or edge matching~\cite{jensen2023jointtrack} needs the bone \emph{outline},
which a plain radiograph does not provide (it presumes manual contouring or a trained edge
detector), whereas our appearance-based correspondence needs no pre-extracted contour and,
being global, supplies the initialization such methods depend on. Given a clean outline, a
silhouette solve is a strong depth refiner, a natural \emph{complement} to ours; we
therefore benchmark against a differentiable-\emph{intensity} renderer (\cref{sec:capture}),
which likewise consumes the raw radiograph. Beyond pose, sharpening the semantic embedding
(\cref{sec:multibone}) into a metric landmark detector is a natural next step.

%% file: sections/06_conclusion.tex
\section{Conclusion}
\label{sec:conclusion}
We present an amortized, subject-agnostic dense 2D--3D correspondence model for single-view
X-ray-to-CT knee registration. One shared-weight model per bone embeds
radiograph pixels and CT points in a common, transmission-aware space, and registers unseen
patients with a global, initialization-free, render-free PnP+RANSAC solve. Trained across a
large cohort, it generalizes to held-out patients with essentially no
seen-to-unseen gap, and the same recipe transfers from the femur to the
patella. Our decomposition shows the residual single-view error is geometric depth, not
correspondence, and is largely resolved by a second view. Against differentiable-rendering
refinement it trades a little accuracy for a far wider capture range at far lower cost.
The embedding is also anatomically semantic: with no bone-level
label it develops a 2D--3D-consistent bone identity, letting a CT segmentation be carried
onto the radiograph, and it identifies landmarks across
patients. On real radiographs registration transfers but is not yet clinical. Closing this
sim-to-real gap is left for future work.